\PassOptionsToPackage{dvipdfmx}{hyperref} 
\documentclass[aps,prl,reprint,groupedaddress, superscriptaddress]{revtex4-2}

\usepackage[dvipdfmx]{graphicx}
\usepackage{color}
\usepackage{float}
\usepackage{natbib}
\usepackage[dvipsnames]{xcolor}
\setcitestyle{numbers}
\setcitestyle{square}
\usepackage{graphicx}
\usepackage{bm}
\usepackage{amssymb}
\usepackage{amsmath}
\usepackage{mathtools}
\usepackage{dcolumn}  
\usepackage[colorlinks, linkcolor=Blue,citecolor=Blue,urlcolor=Blue]{hyperref}

\usepackage[all]{hypcap} 
\usepackage{physics}
\usepackage{enumerate}
\usepackage{braket}       
\usepackage{overpic}
\usepackage{amsfonts}
\usepackage{dsfont}
\usepackage[mathscr]{eucal}
\usepackage{setspace}

\definecolor{mcolorB}{rgb}{0,0,1}
\definecolor{mcolorG}{cmyk}{0.8,0.0,0.5,0.3}
\definecolor{mcolorR}{cmyk}{0,1,0,0}

\newcommand{\+}{\dagger}

\newcommand{\w}{\omega}

\newcommand*{\inp}[1]{\hat{#1}_{\rm in}}
\newcommand*{\out}[1]{\hat{#1}_{\rm out}}

\newcommand{\mcB}[1]{\textcolor{mcolorB}{#1}}

\begin{document}

\title{A Loop-Shaping Approach to Coherent Feedback Control \\
in Cavity Optomechanical Cooling}

\author{Aoi Fujimoto}\affiliation{Department of Mechanical Engineering Informatics, Meiji University, Kawasaki, Kanagawa 214-8571, Japan}
\author{Hiroyuki Ichihara}\affiliation{Department of Mechanical Engineering Informatics, Meiji University, Kawasaki, Kanagawa 214-8571, Japan}
\author{Rina Kanamoto}\affiliation{Department of Physics, Meiji University, Kawasaki, Kanagawa 214-8571, Japan}

\date{\today}

\begin{abstract}
We present a loop-shaping approach to coherent feedback (CF) control. 
By formulating the coupling between a quantum system and its environment in terms of the noise power spectrum, our method enables direct manipulation of the effective dissipation coefficients through spectral shaping. 
A systematic design framework for CF controllers is also developed, in which transfer functions are shaped to realize desired spectral responses. 
Applying this framework to optomechanical sideband cooling, we demonstrate that suppression of the Stokes process and enhancement of the anti-Stokes process can be simultaneously achieved, enabling ground-state cooling even in the unresolved-sideband regime. 
This loop-shaping framework provides an intuitive and general foundation for the design of CF controllers and can be extended to a wide class of quantum systems in which interactions with environments are characterized by noise power spectra.

\end{abstract}

\maketitle

Feedback control of quantum systems is a powerful strategy for achieving diverse objectives \cite{Zhang2017,Wiseman2010} such as rapid purification \cite{Jacobs2003,Combes2006, Combes2010}, quantum error correction \cite{Ahn2002,Gertler2021}, quantum speed limit \cite{Hou2023}, quantum battery \cite{Mitchison2021},  stabilization \cite{Sayrin2011,Vijay2012}, cooling \cite{Rossi2018,Schmid2022,Ghosh2023}, squeezing \cite{Gough2009,Crisafulli2013,Iida2012}, and entanglement generation of qubits and light fields \cite{Shankar2013,Zhou2015}. 
Such feedback is broadly realized in two paradigms:
measurement-based feedback (MF), where the system output is measured and the resulting classical record is processed by a classical controller, and coherent feedback (CF), where a quantum controller is directly coupled to the plant without measurement, thereby preserving coherence and, in various settings, outperforming MF~\cite{Jacobs2014,Yamamoto2014}.
Despite these advantages, systematic design methods for CF controllers remain underdeveloped \cite{Zhang2022}. 
Rigorous approaches such as $H^\infty$ control for linear quantum stochastic systems~\cite{James2008} and coherent linear-quadratic-gaussian (LQG) control with quantum controllers~\cite{Nurdin2009} provide systematic synthesis frameworks but impose physical realizability constraints (such as preservation of canonical commutation relations) that render the design problem nonconvex and computationally demanding.
Consequently, most implementations rely on heuristic, case-specific designs lacking a simple and general design principle.

Feedback control is also applied for quantum control of mechanical oscillators such as levitated nanoparticles~\cite{Li2011, Magrini2021, Tebbenjohanns2021, Kamba2025, Melo2025}, nanomechanical resonators, and moving end mirrors of cavity~\cite{Rossi2018,Schmid2022,Ghosh2023}.
Optomechanical cooling plays a crucial role in preparing mechanical systems near the quantum ground state, which are essential for many optomechanical applications such as quantum transducers~\cite{Barzanjeh2022,Forsch2020,Mirhosseini2020}, atomic clocks and gravitational-wave detection \cite{Aspelmeyer2014}.
In standard sideband cooling \cite{Aspelmeyer2014,WilsonRae2007,Marquardt2007}, ground-state cooling can be achieved only in the resolved-sideband regime, where the mechanical frequency is larger than the cavity linewidth.
However, fundamental optomechanical experiments often employ massive mechanical resonators, which typically place such systems in the unresolved-sideband regime \cite{Guo2022}, where the cavity linewidth is larger than the mechanical frequency.
This limitation has become a key obstacle to realizing many quantum applications of optomechanical systems.
Both MF \cite{Rossi2018,Guo2023,Wang2023} and CF \cite{Huang2019,Guo2022,Ernzer2023} schemes have been extensively explored for cooling mechanical motion.

In this Letter, we propose a loop-shaping framework for CF controller design, providing a systematic approach to shape quantum noise spectra and control system dynamics.
In this framework, we can determine controller parameters intuitively before constructing the closed-loop model, offering a transparent alternative to conventional trial-and-error tuning. The aim of this work is two-fold:
(i) to establish a loop-shaping framework for coherent feedback control; and
(ii) to demonstrate how it enables ground-state cooling of a mechanical oscillator in the unresolved-sideband regime.

\begin{figure}[t]
  \centering
  \includegraphics[width=0.9\linewidth]{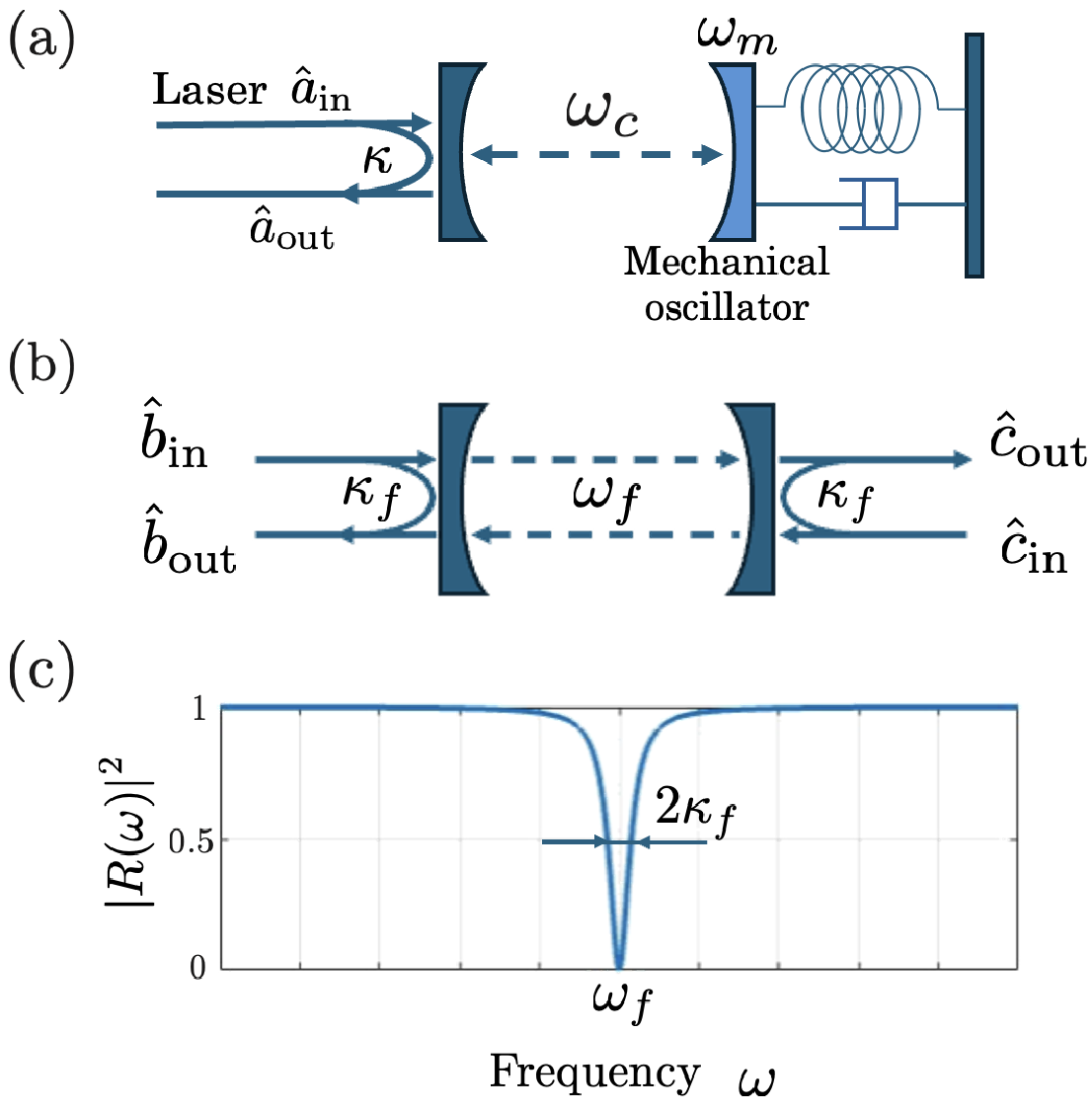}
  \caption{ 
(a) Canonical optomechanical setup, where the mechanical oscillator with frequency $\omega_m$ couples to the intracavity field at frequency $\omega_c$ via radiation pressure.
(b) Double-sided cavity acting as the CF controller, characterized by equal mirror decay rates $\kappa_f$ and cavity resonance $\w_f$.
(c) Frequency response of the reflection coefficient $|R(\omega)|^2$ of the double-sided cavity.
At the cavity resonance frequency $\w_f$, the reflectance vanishes as $|R(\w_f)|^2 = 0$.}
  \label{fig:1}
\end{figure}

Figure~\ref{fig:1}(a) shows a canonical optomechanical setup, where 
the displacement of the mechanical oscillator, $\hat{X}=x_{\rm ZPF}(\hat{b}_m^\dagger+\hat{b}_m)$, is coupled to the linearized radiation-pressure force of the intracavity field,
$\hat{F}_{\rm rad}=\hbar g_0(\hat{a}_c^\dagger+\hat{a}_c)$, through the interaction Hamiltonian $\hat{H}_{\rm int}=-\hat{F}_{\rm rad}\hat{X}$.
Here, $x_{\rm ZPF}$ is the zero-point fluctuation amplitude, $g_0$ is the linearized optomechanical coupling, and $\hat{a}_c$ and $\hat{b}_m$ are the annihilation operators of the intracavity field and the mechanical mode, respectively.

The laser cooling of the mechanical oscillator arises from an imbalance between the Stokes and anti-Stokes scattering processes, which correspond to phonon creation and annihilation, respectively~\cite{Clerk2010}. The mean phonon number evolves according to the Stokes rate $A_+$ and anti-Stokes rate $A_-$, and the minimum achievable occupation is given by $\bar{n}_{\rm min}=A_+/(A_--A_+)$.
The scattering rates $A_\pm$ are determined by the noise power spectral density of the radiation-pressure $S_{FF}(\omega)$ as 
\begin{equation}
  A_\pm = \frac{x_{\rm ZPF}^2}{\hbar^2}S_{FF}(\w=\mp \w_m),\label{eq:1}
\end{equation}
where $\omega_m$ is the mechanical frequency.

For conventional sideband cooling without feedback, these rates take the form
$A_\pm = g^2 \kappa /[(\kappa/2)^2 + (\Delta \mp \omega_m)^2]$,
where $g=g_0 x_{\rm ZPF}$ and $\Delta=\omega_L-\omega_c$ is the detuning of the drive-laser frequency $\omega_L$ from the cavity resonance $\omega_c$.
Here $\kappa$ denotes the cavity decay rate through the external coupling port, assuming negligible internal losses.
For red detuning ($\Delta < 0$), the anti-Stokes process dominates ($A_+<A_-$), leading to cooling.
The optimal cooling efficiency is achieved at $\Delta=-\omega_m$, where the optimal damping rate, $A_- - A_+$, is maximized.
In the resolved-sideband limit ($\kappa \ll \omega_m$), this yields $\bar{n}_{\min}<1$ and enables ground-state cooling.
In contrast, in the unresolved-sideband limit ($\kappa \gg \omega_m$), the Stokes and anti-Stokes sidebands are not spectrally resolved, residual Stokes scattering yields $\bar{n}_{\min} > 1$, preventing ground-state cooling in standard approach.

To overcome this limitation, we employ CF control based on a loop-shaping approach.
In this framework, the system dynamics are modeled by transfer functions~\cite{Yanagisawa2003a,Yanagisawa2003b}, enabling frequency-domain analysis and controller design in a systematic and physically transparent manner.
This allows the radiation-pressure noise to be shaped so that the Stokes rate $A_+$, which contributes to heating, is suppressed, while the anti-Stokes rate $A_-$, responsible for cooling, is enhanced.

We begin by focusing on the suppression of the Stokes process, which is responsible for heating. As shown in Eq.~\eqref{eq:1}, the Stokes rate $A_+$ is proportional to the quantum noise spectral component at frequency $-\omega_m$ in the rotating frame of the drive-laser frequency. We therefore aim to reduce it by employing a controller that attenuates this component in the system response. 
The controller should suppress the spectral component at $-\omega_m$ while leaving that at $+\omega_m$ essentially unaffected, since the latter contributes to the anti-Stokes process and thus to cooling. Suitable filter types include high-pass filters, notch filters, and band-pass filters. In the following, we illustrate this idea with concrete realizations based on a two-sided cavity, which can be configured either as a notch filter, which strongly attenuates a narrow band around a specific frequency, or as a band-pass filter, which transmits only a selected frequency band, depending on how it is interconnected with the system.

\begin{figure}[t]
  \centering
  \includegraphics[width=1\linewidth]{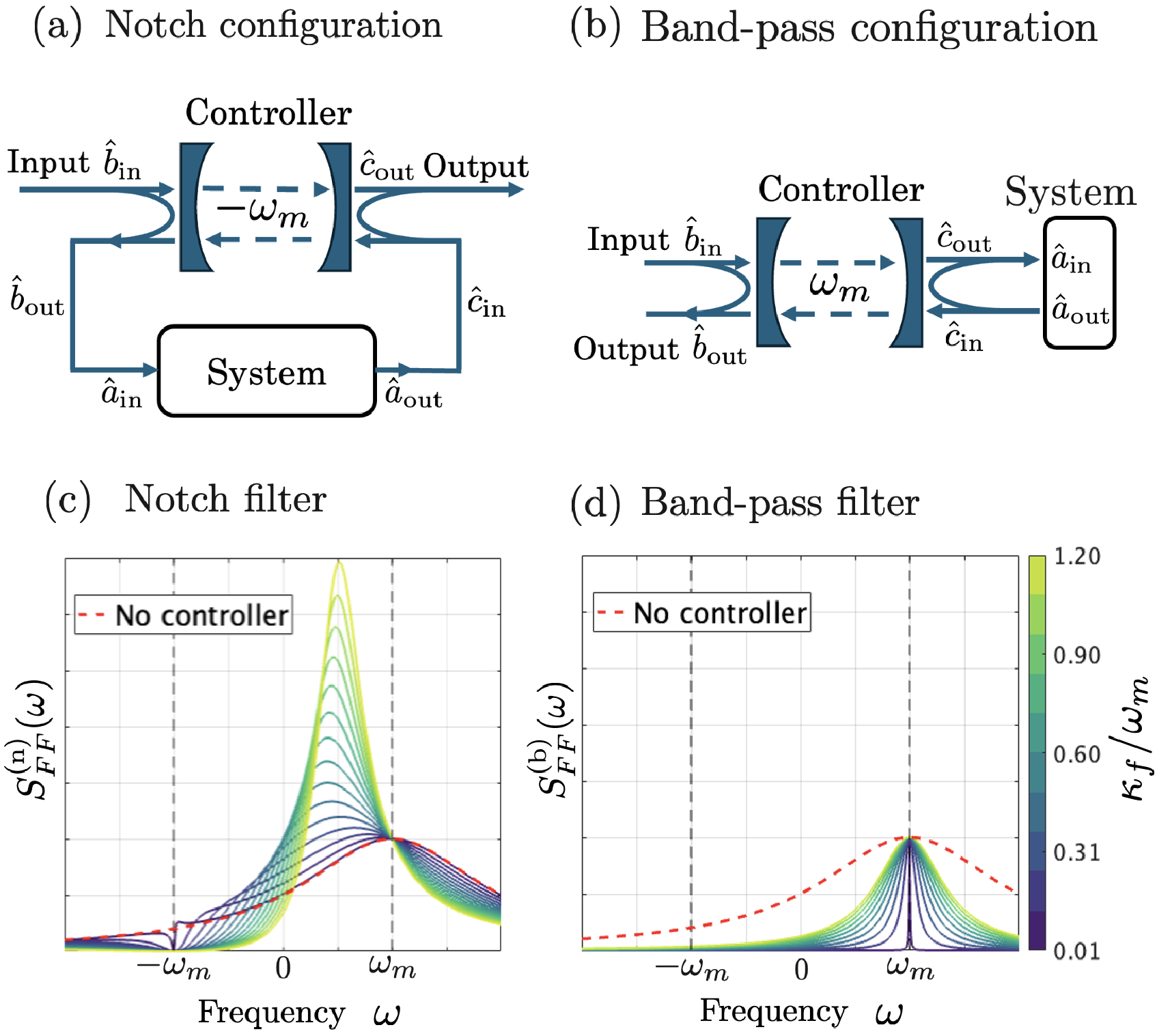}
  \caption{
(a,b) CF loop configurations operating as a notch filter (a) and a band-pass filter (b).
Both configurations suppress the Stokes sideband at $-\omega_m$: the notch filter directly blocks this frequency component, while the band-pass filter achieves suppression by transmitting only the anti-Stokes sideband at $+\omega_m$.
(c,d) Radiation-pressure noise spectra corresponding to (a) and (b), denoted as $S^{(\rm n)}_{FF}(\omega)$ and $S^{(\rm b)}_{FF}(\omega)$, respectively, calculated at $\Delta=-\omega_m$.
In both spectra, the red dashed curves represent the case without the CF controller.
  }
  \label{fig:2}
\end{figure}

As a CF controller, we consider a double-sided cavity, shown schematically in Fig.~\ref{fig:1}(b), which is a two-input, two-output system with transmission through both mirrors. The intracavity mode is denoted by the annihilation operator $\hat{a}_f$, with resonance frequency $\w_f$. The cavity is driven by a laser at frequency $\omega_L$, and the detuning is defined as $\Delta_f=\w_L-\w_f$. Again, we assume no internal loss, and take both mirrors to have identical decay rates $\kappa_f$. The input and output field operators on the left (right) ports are denoted by $\inp{b}$ and $\out{b}$ ($\inp{c}$ and $\out{c}$), respectively.
In the rotating frame at the drive-laser frequency $\w_L$, the dynamics are governed by the Langevin equation
\begin{equation}
  \dot{\hat{a}}_f(t)=\left(i\Delta_f-\kappa_f\right)\hat{a}_f(t)-\sqrt{\kappa_f}\inp{b}(t)-\sqrt{\kappa_f}\inp{c}(t)\notag
\end{equation}
together with the input-output relations
\begin{align}
    \out{b}(t) &= \sqrt{\kappa_f}\hat{a}_f(t) + \inp{b}(t),\notag\\
  \out{c}(t) &= \sqrt{\kappa_f}\hat{a}_f(t) + \inp{c}(t)\notag.
\end{align}
Fourier transforming yields the transfer relation
\begin{align}
  \begin{bmatrix}
    \out{b}(\w)\\
    \out{c}(\w)
  \end{bmatrix}&=\begin{bmatrix}
    R(\w) & T(\w)\\
    T(\w) & R(\w)
  \end{bmatrix}
  \begin{bmatrix}
    \inp{b}(\w)\\
    \inp{c}(\w)
  \end{bmatrix},\notag
\end{align}
with reflection and transmission coefficients
\begin{equation}
  R(\w)  =\frac{i(\w+\Delta_f)}{i(\w+\Delta_f)-\kappa_f},~~~~T(\w)=\frac{\kappa_f}{i(\w+\Delta_f)-\kappa_f}\notag\mcB{,}
\end{equation}
satisfying $|R(\omega)|^2+|T(\omega)|^2=1$. 
As shown by the frequency response of $R(\omega)$ in Fig.~\ref{fig:1}(c), the double-sided cavity serving as the CF controller reflects quantum noise at frequencies away from its resonance $\w_f$ and fully transmits the component exactly at $\w_f$ through the controller.
By connecting the double-sided cavity to the system as illustrated in Fig.~\ref{fig:2}(a) and tuning its resonance to $\w_f=\omega_L-\omega_m$ (equivalently, $\Delta_f=\omega_m$), the control cavity blocks the spectral component at $-\omega_m$.
Under this condition, the CF controller acts as a notch filter that suppresses the Stokes process, as in our previous work~\cite{Fujimoto2025}.
In this feedback configuration, the radiation-pressure noise spectrum of the system under CF control is given by 
$S^{(\rm n)}_{FF}(\w)=\hbar^2g^2|\chi^{(\rm n)}(\w)|^2/x_{\rm ZPF}^2$, 
where
\begin{equation}
  \chi^{(\rm n)}(\w)=\frac{\chi(\w)R(\w)}{1-(\sqrt{\kappa}\chi(\w)+1)T(\w)}\notag
\end{equation}
 represents the transfer function from the input field $\inp{b}$ to the intracavity field $\hat{a}_c$ in Fig.~\ref{fig:2}(a). Here, $\chi(\omega)=\sqrt{\kappa}/[i(\Delta+\omega)-\kappa/2]$ denotes the transfer function from the input field $\hat{a}_{\rm in}$ to $\hat{a}_c$ in Fig.~\ref{fig:1}(a), also referred to as the optical susceptibility of the cavity.
The corresponding Stokes and anti-Stokes rates under $\Delta = -\w_m$ are obtained from Eq.~(\ref{eq:1}) as
\begin{align}
  A_+^{(\rm n)} &= \frac{x_{\rm ZPF}^2}{\hbar^2}S^{(\rm n)}_{FF}(- \w_m) = 0\label{eq:2},\\
  A_-^{(\rm n)} &= \frac{x_{\rm ZPF}^2}{\hbar^2}S^{(\rm n)}_{FF}(+ \w_m) = \frac{4g^2}{\kappa}.\notag
\end{align}
Figure~\ref{fig:2}(c) shows the spectrum $S^{(\rm n)}_{FF}(\omega)$ for the case of notch configuration.
As seen in Fig.~\ref{fig:2}(c) and Eq.~(\ref{eq:2}), the spectral component at $-\omega_m$ is completely suppressed by the CF controller, confirming the full elimination of the Stokes process.
In contrast, the component at $+\omega_m$ remains unchanged from the uncontrolled case, and thus the anti-Stokes rate $A_-^{(\rm n)}$ retains the same value as before control. 
It follows that the minimum phonon number is always $\bar{n}_{\rm min}=0$, indicating that ground-state cooling can in principle be achieved independent of the cavity decay rate~$\kappa$.

The double-sided cavity can also operate as a band-pass filter when connected to the system as illustrated in Fig.~\ref{fig:2}(b). To suppress the Stokes process using this configuration, the controller must transmit only the spectral component at $+\omega_m$. This is achieved by tuning the controller cavity resonance to $\w_f=\omega_L+\omega_m$, or equivalently $\Delta_f=-\omega_m$.
In this case, the radiation-pressure noise spectrum of the system under CF control is given by 
$S^{(\rm b)}_{FF}(\w)=\hbar^2g^2|\chi^{(\rm b)}(\w)|^2/x_{\rm ZPF}^2$, 
where
\begin{equation}
  \chi^{(\rm b)}(\w)=\frac{\chi(\w)T(\w)}{1-(\sqrt{\kappa}\chi(\w)+1)R(\w)}\notag
\end{equation}
represents the transfer function from the input field $\inp{b}$ to the intracavity field $\hat{a}_c$ in Fig.~\ref{fig:2}(b).
The corresponding Stokes and anti-Stokes rates at $\Delta = -\w_m$ are obtained as
\begin{align}
  A_+^{(\rm b)} &= \frac{x_{\rm ZPF}^2}{\hbar^2}S^{(\rm b)}_{FF}(- \w_m) = \frac{g^2 \kappa }{(\kappa/2)^2 + 4(\kappa/\kappa_f+1)^2\w_m^2},\notag\\
  A_-^{(\rm b)} &= \frac{x_{\rm ZPF}^2}{\hbar^2}S^{(\rm b)}_{FF}(+ \w_m) = \frac{4g^2}{\kappa}\notag.
\end{align}
Figure~\ref{fig:2}(d) shows the spectrum $S^{(\rm b)}_{FF}(\omega)$ for the case of band-pass configuration.
In contrast to the notch-filter case, the band-pass configuration does not completely suppress the Stokes process.
The degree of suppression depends on the decay rate $\kappa_f$ of control cavity, which determines the bandwidth of the band-pass filter: smaller $\kappa_f$ yields stronger suppression of the Stokes rate.
In the limit $\kappa_f \to \infty$, the Stokes rate becomes identical to that of the uncontrolled system.
The anti-Stokes rate $A_-^{(\rm b)}$ remains unchanged from both the notch-filter case and the uncontrolled case.
When $\kappa\kappa_f / [4(\kappa+\kappa_f)] < \omega_m$, ground-state cooling can be achieved using this band-pass configuration.

As demonstrated by the two examples of the notch and band-pass filters,
the controller parameters can be determined a priori from the desired frequency-domain response, without explicitly modeling the closed-loop system.
Unlike conventional approaches that tune controller parameters by analyzing the closed-loop dynamics,
this method significantly reduces computational overhead and enhances physical transparency in the design process.

\begin{figure}[t]
  \centering
  \includegraphics[width=0.9\linewidth]{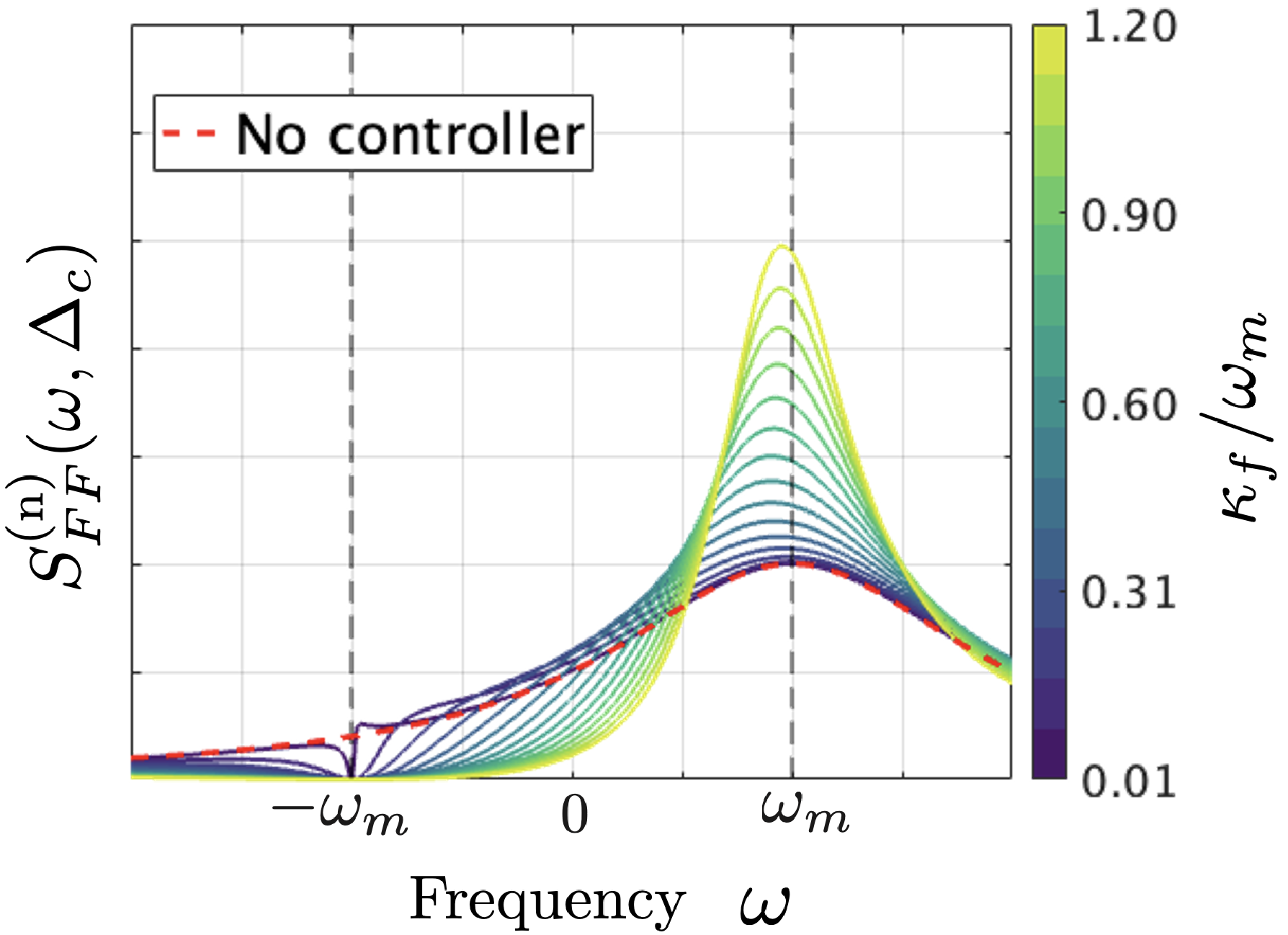}
  \caption{
    Radiation-pressure noise spectrum $S^{(\rm n)}_{FF}(\omega,\Delta_c)$ of the system under the CF control with the notch-filter configuration shown in Fig.~2(a).
At the optimal detuning $\Delta=\Delta_c$, the anti-Stokes process is enhanced by a factor of $1+(\kappa_f/\omega_m)^2$, while the Stokes component at $-\omega_m$ remains completely suppressed.
The red dashed curve represents the spectrum under $\Delta=-\omega_m$ without the CF controller.
  }
  \label{fig:3}
\end{figure}

We now focus on enhancing the anti-Stokes process.
While suppressing the Stokes process is sufficient to achieve ground-state cooling in the unresolved-sideband regime, enhancing the anti-Stokes process further improves the cooling efficiency and robustness against inevitable disturbances.
To enhance the anti-Stokes process, one could in principle employ gain elements to amplify the spectral component at frequency $+\omega_m$ in the system response. However, here we take a simpler approach using the same notch-filter configuration, exploiting the additional resonance mode created by the feedback loop to enhance the anti-Stokes response.
As shown in Fig.~\ref{fig:2}(c), increasing the decay rate $\kappa_f$ of the control cavity leads to an additional resonance mode in the feedback loop, resulting in a higher peak in the optical response.
However, this peak does not coincide with the anti-Stokes sideband at frequency $+\omega_m$, and therefore does not directly enhance the anti-Stokes rate.
To maximize the anti-Stokes process under CF control, we determine the detuning $\Delta_c$ that maximizes $A_-^{(\rm n)}(\Delta)$, given by
\begin{align}
  \Delta_c =\underset{\Delta} {\operatorname{argmax}}~A_-^{(\rm n)}(\Delta) =- \w_m - \frac{\w_m\kappa_f\kappa}{2(\w_m^2+\kappa_f^2)}\notag.
\end{align}
The corresponding Stokes and anti-Stokes rates evaluated at $\Delta=\Delta_c$ are
\begin{align}
  A_+^{(\rm n)}(\Delta_c) = 0,~~~A_-^{(\rm n)}(\Delta_c)= \frac{4g^2}{\kappa}\left\{1+\left(\frac{\kappa_f}{\w_m}\right)^2\right\}\label{eq:3}.
\end{align}
Figure~\ref{fig:3} shows the spectrum $S^{(\rm n)}_{FF}(\w)$ at $\Delta=\Delta_c$.
As seen from Eq.~(\ref{eq:3}) and Fig.~\ref{fig:3}, choosing the optimal detuning $\Delta_c$ enhances the anti-Stokes rate by a factor of $1+\kappa_f^2/\omega_m^2$ compared with the conventional approach, thereby demonstrating that the anti-Stokes process is indeed promoted under CF control.

While the present results demonstrate idealized performance, achieving $\bar{n}_{\mathrm{min}} = 0$ in experiments is challenging due to several practical imperfections.
First, as in conventional schemes, internal cavity losses and propagation losses in the coupling waveguides degrade the filter performance.
In addition, any imbalance in the decay rates of the two mirrors in the double-sided cavity prevents complete suppression of the Stokes process as described by Eqs.~(\ref{eq:2}) and (\ref{eq:3}).
Nevertheless, these imperfections mainly reduce the filtering efficiency without altering the qualitative behavior, so the proposed method should still outperform conventional sideband cooling in realistic conditions.
Feedback delay is another important factor, as it introduces frequency-dependent phase shifts that can significantly alter the effective frequency response. Therefore, the delay and the controller parameters must be tuned jointly. Time-delayed coherent feedback has been extensively studied~\cite{Ernzer2023,Grimsmo2014,Grimsmo2015}, and appropriate tuning of the delay may even enhance cooling performance under certain conditions. Combining our loop-shaping framework with such delay-engineered coherent feedback may thus enable even more efficient cooling strategies. 
Although the present work focuses on passive CF controllers, employing active elements such as degenerate parametric amplifiers could further improve the cooling efficiency, at the cost of increased susceptibility to excess noise.

The proposed framework is applicable to a broad class of open quantum systems weakly coupled with environments given by the interaction Hamiltonian $\hat{H}_{\rm int} = \sum_k\Bigl[\hat{L}_k\hat{E}_k^\+ + \hat{L}^\+_k\hat{E}_k\Bigl]$,
where $\hat{L}_k$ and $\hat{E}_k$ denote the system and environmental operators, respectively.
Under the Born--Markov and rotating--wave approximations, the reduced dynamics of the system is given by the Lindblad master equation
\begin{equation}
  \frac{d\hat{\rho}}{dt}=-\frac{i}{\hbar}[\hat{H}_{\rm sys},\hat{\rho}]+\sum_k\left(\gamma_{k}^-\mathcal{D}[\hat{L}_k]\hat{\rho}+\gamma_{k}^+\mathcal{D}[\hat{L}^\+_k]\hat{\rho}\right), \notag
\end{equation}
where $\mathcal{D}[\hat{L}]\hat{\rho}\equiv \hat{L}\hat{\rho}\hat{L}^\+ - \{\hat{L}^\+\hat{L},\hat{\rho}\}/2$.
The coefficients $\gamma_k^\pm$ are determined by the environmental spectrum \cite{Breuer2002,Clerk2010}, and in the optomechanical case they correspond directly to the Stokes and anti-Stokes rates $A_\pm$ introduced above.
Coherent feedback enables the shaping of the environmental spectrum, thereby allowing direct control of the effective dissipation coefficients $\gamma_k^\pm$.
This frequency-domain perspective provides a versatile framework for quantum dissipation engineering in CF control, extending beyond cooling to include oscillation phenomena and entanglement generation.

In summary, we have developed a loop-shaping framework for coherent feedback control, providing a systematic and intuitive frequency-domain approach to shaping quantum-noise spectra and controlling system dynamics.
Applying this approach to optomechanical sideband cooling, we demonstrated that ground-state cooling can be achieved in the unresolved-sideband regime with enhanced cooling efficiency, confirming the practical utility of the proposed framework.

We thank M. Bhattachayra for fruitful discussions. This work was supported by JSPS KAKENHI (Grant No. JP25K07190 and No. JP23K03915) and JST ERATO (Grant No. JPMJER2302).

\bibliographystyle{apsrev4-2}
\bibliography{references}

\end{document}